# Time-varying Vine Copula model based on R-Vine structure and its application in financial risk research

XueZeng Yu[a]

[a]Siyang Kangaroo Mother Childcare Co., Ltd.,Jiangsu,China.          September 14, 2025

## ABSTRACT

The time-varying Vine Copula model has become a new direction in the Vine Copula class of models due to its time-varying structural parameters. We have observed that the Vine structures of the time-varying Vine Copula model currently used in economics and business research are C-Vine and D-Vine. These two structures are simpler than the R-Vine structure in modeling but will lose more details. Although truncation and simplification of the Vine structure are necessary when the number of variables is large, the number of variables in economics and business research is often small. Therefore, the R-Vine structure is definitely more suitable for constructing time-varying Vine Copula for economic research. Therefore, this paper uses the GAS (Generalized Autoregressive Score) model to dynamically parameterize the R-Vine structure to construct a time-varying Vine Copula model. The application of this model to the study of liquidity risks between China and Southeast Asian countries, including during the pandemic period, reveals that the time-varying model based on the R-Vine structure not only achieves better statistical test results but also better reflects economic and even political realities compared to the other two structural time-varying models.

**KEYWORDS:** R-Vine;Time-Varying;Vine Copula.

CONTACT   XueZeng Yu    Email: yu_tom@alu.gxu.edu.cn    Siyang Kangaroo Mother Childcare Co.,Ltd., Jiangsu,China.

## 1. Introduction

The Copula model has been widely applied in depicting the interdependence among variable data, which is widely used in the field of financial research for high-frequency data analysis, such as Weiss and Supper(2013),Amin et al.(2021), Abakah et al.(2021). Based on the multivariate copula, Bedford and Cooke (2001)first proposed the Vine theory. According to this theory, the multivariate copula function can be decomposed into a multi-layer acyclic tree structure consisting of binary copula and binary conditional copula function, thereby obtaining richer linkage information. Matthias et al.(2009) compared the modeling effects of Multivariate Copula and Vine Copola on financial return data and found that Vine Copula had better modeling effects. Subsequently, Czado et al.(2009),Patton(2006) and Brechmann et al.(2013)further promoted the development of Vine Copula models. Aas et al.(2009)proposed the C-Vine and D-Vine structures, which are specialized Vine structures. When the number of variables is large, this helps to simplify the modeling of the Vine Copula model. Compared with Bayesian networks/DAG models (Pearl(1988),Lauritzen(1996),Koller and Friedman(2009)), the dozens of binary Copulas that currently characterize the nonlinear dependency relationship between two variables make the Vine Copula model have advantages in characterizing richer linkage characteristics between variables and predicting functions.

Financial connections change in real-time due to the influence of external environments, so Patton(2001)first proposed a time-varying binary Copula model based on the ARMA (1,10) framework. Since Heinen and Valdesogo(2009) first constructed the time-varying Vine Copula model with the C-Vine structure, many scholars have subsequently constructed time-varying Vine-Copula models with the C-Vine structure and the D-Vine structure. Among them, there are some papers that construct time-varying models based on D-Vine structure (such as XingYu et al.,2020; QunWei et al.,2022;Manner et al.,2019;Rui and Min,2021). These papers all use the Patton method to dynamicize the binary Pair-Copula parameters. Other literature(such as Anubha and Aparna,2019;Jinguo and Ruqiao,2015; Yingwei and Jie,2022)constructed a time-varying model based on the C-Vine structure, with most of the literature using the Patton method. Jinguo and Ruqiao (2015) and YingWei and Jie(2022)used the GAS model to dynamically transform the binary Pair Copula and construct a time-varying Vine Copula model. The above literature has demonstrated that time-varying models have better fitting performance when compared with static models. From the above existing literature on time-varying Vine Copula models, we can find that some literature, such as Rui and Min(2021), Manner et al.(2019), and Anubha and Aparna (2019), directly choose C-Vine or D-Vine in the selection of Vine structure, without considering other types of Vine structures; other literature can only choose C-Vine structure and D-Vine structure for modeling due to the small number of variables. For example, when there are three variables, the C-Vine structure is the same as the D-Vine structure and there is only one structure. However, when there are more than five variables, the structure between the variables is often not the C-Vine structure or the D-Vine structure. The C-Vine structure stipulates that the connection relationship between variables is radial, and the D-Vine structure stipulates that the connection relationship between variables is completely parallel. The R-Vine structure does not have these rules. Therefore, choosing the

R-Vine structure will increase the complexity of modeling and numerical solution. C-Vine and D-Vine are both special R-Vine structures in a broad sense, Dißmann et al.(2013)deduced that when there is an R-Vine structure between variables instead of a C-Vine/D-Vine structure, the modeling performance of the Vine Copula model based on this R-Vine structure will be better than the previous two; Wei et al. (2014) also pointed out that in high-dimensional cases, the R-Vine structure shows more realistic linkage relationships than the other two structures. The number of variables in the above literature is almost always less than 10. The simplification effect of the C-Vine and D-Vine structures can only be effective when the number of variables is much greater than 10. Therefore, this article believes that, in general, the R-Vine structure should be Considered as the preferred model in economics and business research.

In order to further improve the modeling performance of the time-varying Vine Copula model, this article selects the GAS (Generalized Autoregressive Score) model to perform dynamic binary Pair-Copula parameters, which can overcome the problems of traditional ARMA models in terms of parameter updating mechanism. Therefore, this paper constructs a new time-varying Vine Copula model based on the R-Vine structure and the GAS-Copula model, and also constructs a model based on the C-Vine and D-Vine structures using GAS-Copula for application comparison. Based on the proposed new model, this article examines liquidity risk between China and major Southeast Asian countries from 2018 to 2023. The results show that the time-varying Vine Copula model constructed based on the R-Vine structure not only outperforms the other two time-varying Vine Copula models in various statistical tests but also more realistically reflects the economic and political relations between countries in the region. In addition, the article also found that the effect of residual modeling on the original data will affect the choice of Vine structure.

The remainder of the article is organized as follows: Part II analyzes ARMA-GARCH models for modeling marginally distributed data, introduces Vine Copula and time-varying Copula models, and discusses the construction and parameter estimation of the time-varying R-Vine Copula model. Part III applies modeling to data from China and major Southeast Asian countries, constructs time-varying D-Vine Copula and C-Vine Copula models, and statistically tests the modeling results of these three models on the same data, comparing their ability to explain economic and political factors. The final part summarizes the article's findings and provides an outlook.

## 2. Time-varying R-Vine Copula Model

The variables of the copula function must be uniformly distributed between [0,1]. Therefore, first, use a time series model such as the ARFIMA-GARCH model to extract various features from the raw data to obtain standardized residuals. Then, use the probability integral transformation method to convert these residuals into uniformly distributed data between [0,1]. Only then can the copula model be constructed and the next steps can be performed.

### 2.1. ARFIMA-GARCH Model

Time series data often have characteristics such as heteroskedasticity, volatility clustering and non-stationarity. Since its birth in the 1980s, the ARMA-GARCH model has been maturely applied to the modeling of various time series data. Currently, most of the literature that uses Copula models for research uses this model to process raw data. However, the ARMA-GARCH model is difficult to fit the long memory and non-stationarity that often exist in time series data.

Therefore, this article draws on the modeling methods of Brechmann(2018), a scholar who made outstanding contributions to the basic theory of Vine-Copula, and Rehman(2020), Singh et al.(2021), and uses the ARFIMA-GARCH model with fractional difference operation to extract the residual sequence of the original data. The ARFIMA(p,d,q)-GARCH(a,b) model can effectively filter out long memory and non-stationary characteristics so that the data meets the requirements of the Copula model. For one-dimensional data, the model is defined by three equations:

$$\text{Mean equation: } \Phi(L)(1-L)^d r_t = \Theta(L)\varepsilon_t ;$$

$$\text{Conditional heteroskedasticity equation: } \sigma_t^2 = \omega + \sum_{i=1}^{p} \alpha_i \varepsilon_{t-i}^2 + \sum_{j=1}^{q} \beta_j \sigma_{t-j}^2 ; \quad (1)$$

$$\text{Volatility equation: } \varepsilon_t = \sigma_t z_t; \ z_t \sim N(0, 1).$$

In the above equation, L in the mean equation is the lag operator, $(1-L)^d$ is the fractional d-order difference operation, d represents the fractional difference and -0.5<d<0.5, $\Phi(L)$ and $\Theta(L)$ are the p-order and q-order stationary autoregressive operators and reversible moving average operators respectively, $\Phi(L) = 1 - \emptyset_1 L - \emptyset_2 L^2 - \ldots \ldots - \emptyset_p L^p$, $\Theta(L) = 1 - \Theta_1 L - \Theta_2 L^2 - \ldots \ldots - \Theta_q L^q$. The mean equation selectively uses fractional difference operations based on whether the data is a stationary series and has long memory characteristics. If fractional differencing is not required, the mean equation becomes:

$$r_t = \mu + \sum_{i=1}^{p} \alpha_i r_{t-i} + \sum_{j=1}^{q} \beta_j \varepsilon_{t-j} + \varepsilon_t. \quad (2)$$

In order to facilitate the solution of the overall model, the paper uses the ARFIMA(p,d,q)-GARCH(a,b)-sstd model to model the raw data. Finally, in order to reflect the comparative effect of the model after adding fractional difference operation, the article uses the ARMA-GARCH model to model the raw data in the empirical application part, and then models the obtained data with the time-varying R-Vine Copula model.

2.2. Vine Copula model

The copula model properties are defined by Sklar (1959), Let the joint distribution function of n-dimensional random variables $x_1, x_2,..., x_n$ be $F(x_1, x_2,..., x_n)$, $f(x_1, x_2,..., x_n)$ is its probability density function, $F_1(\cdot), F_2(\cdot),..., F_n(\cdot)$ is its continuous marginal probability distribution function. Then there is a unique n-dimensional distribution function $C(u_1, u_2,..., u_n) = C(F_1(\cdot), F_2(\cdot),..., F_n(\cdot)) = F(x_1, x_2,..., x_n)$; Therefore $x_n = F_n^{-1}(u_n)$, $u_n$ follows a uniform distribution on [0,1]. $C(\cdot)$ is the Copula distribution function, and there is also a conditional Copula distribution function accordingly. The Vine theory decomposes the joint distribution probability density function of the residuals of multiple variable data into a form consisting of the multiplication of a binary copula probability density function and a binary conditional copula probability density function to obtain richer linkage details between variables. The connection structure between these binary/binary conditional copula probability density functions can be represented by a tree structure, and the formation process of these tree structures follows certain rules. For each layer of tree $T_i$, it consists of nodes $N_i$ and edges $E_i$. For the regular Vine structure, its mathematical structure can be expressed by the following n-dimensional probability density function expression:

$$f(x_1,..., x_n) = \prod_{d=1}^{n} f_d(x_d) \prod_{j=n-1}^{1} \prod_{i=n}^{j+1} C_{m_{j,j}, m_{i,j}|m_{i+1,j},...,m_{n,j}}(F_{m_{j,j}|m_{i+1,j},...,m_{n,j}}, F_{m_{i,j}|m_{i+1,j},...,m_{n,j}}). \quad (3)$$

$F(\cdot)$ is the marginal distribution function, and $m_{i,j}$ is the element in the R-Vine matrix(Oswaldo,2010). $c_m(\cdot)$ is the binary/binary conditional copula probability density function, $f_d(\cdot)$ is the marginal probability density function, $F_{x|v}$ is the conditional marginal distribution function, $v$ is the condition set. The solution process can be expressed by the

following equation:

$$F(x|v) = \frac{\partial C_{xv_j|V_{-j}}(F_x(x|V_{-j}), F_v(v_j|V_{-j}))}{\partial F_v(v_j|V_{-j})}. \quad (4)$$

The capital letter V represents the condition set, $v_j \in V$ and $V_{-j}$ represents the remainder of V after removing $v_j$. The process of solving these conditional marginal distribution functions is an important part of the iterative solution between different tree layers of the Vine structure. For n variables, their Vine structure is not unique. The following is an example of a 5-dimensional regular vine (R-vine), and its decomp -osition process and structure are as follows:

$$f_{12345}(x_1, x_2, x_3, x_4, x_5) = f_1(x_1) \cdot f_2(x_2) \cdot f_3(x_3) \cdot f_4(x_4) \cdot f_5(x_5) \cdot$$

$$c_{12}(F_1(x_1), F_2(x_2)) \cdot c_{23}(F_2(x_2), F_3(x_3)) \cdot c_{34}(F_3(x_3), F_4(x_4)) \cdot c_{35}(F_3(x_3), F_5(x_5)) \cdot$$

$$c_{13|2}(F_{1|2}(x_1|x_2), F_{3|2}(x_3|x_2)) \cdot c_{42|3}(F_{2|3}(x_2|x_3), F_{4|3}(x_4|x_3)) \cdot c_{45|3}(F_{4|3}(x_4|x_3), F_{5|3}(x_5|x_3)) \cdot$$

$$c_{14|23}(F_{4|23}(x_4|x_2, x_3), F_{1|23}(x_1|x_2, x_3)) \cdot c_{25|34}(F_{5|34}(x_5|x_3, x_4), F_{2|34}(x_2|x_3, x_4)) \cdot$$

$$c_{15|234}(F_{5|234}(x_5|x_2, x_3, x_4), F_{1|234}(x_1|x_2, x_3, x_4)). \quad (5)$$

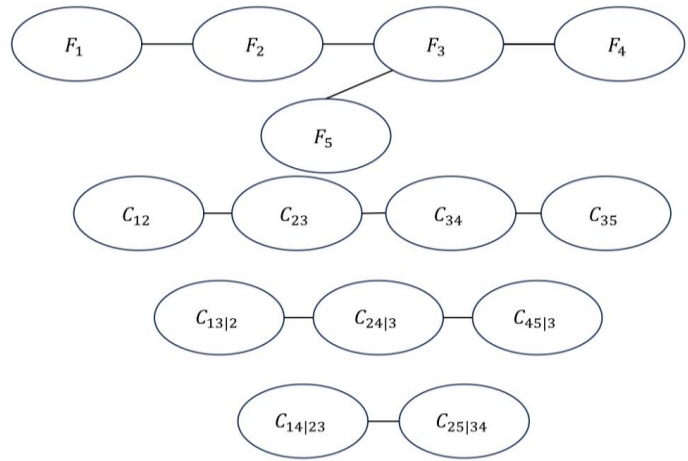

Figure 1. 5-variable R-Vine tree structure.

The more special types of Vine structures are D-Vine and C-Vine. Their tree structure construction process is simpler or more regular than that of R-Vine. For the D-Vine structure, each layer of the tree is a parallel distribution structure. When the first layer of the tree structure is determined, the structure of the other layers of the tree is also determined. Take the first layer of the 5-dimensional variable tree structure as an example:

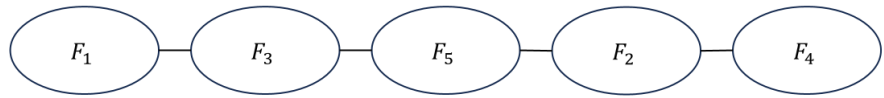

Figure 2 The first-level tree structure of the 5-variable D-Vine.

For the C-Vine structure, each layer of the tree is a central radial distribution structure. Take the first layer tree structure diagram of the 5-dimensional variable as an example:

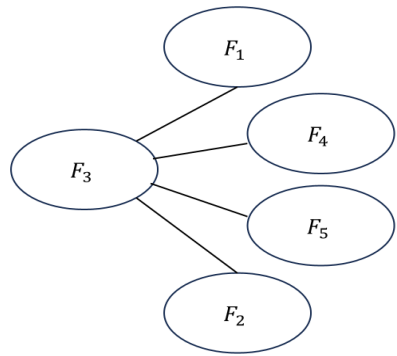

Figure 3. The first-level tree structure of the 5-variable C-Vine.

2.3. Time-Varying Copula Model

Patton(2001) first studied the time-varying binary Copula model, where the interdependence between high-frequency data varies over time. Therefore, time series modeling of the dependent structural parameters of the Copula function was used to characterize its time-varying nature. Taking the Guassian-Copula function as an example, the model uses a process similar to ARMA(1,10) to construct a time-varying binary Guassian-Copula function:

$$C(x,y;\rho)=\int_{-\infty}^{\Phi^{-1}(x)} \int_{-\infty}^{\Phi^{-1}(y)} \frac{1}{2\pi\sqrt{1-\rho^2}} \exp\left\{\frac{-(r^2+s^2-2\rho rs)}{2(1-\rho^2)}\right\} drds \qquad (6)$$

Where $\Phi^{-1}(\cdot)$ is the inverse function of the standard normal distribution function. $\rho \in $ (-

1,1) is a related parameter and is the object of time series modeling. Further modeling the parameter ρ using a process similar to ARMA (1,10):

$$\rho_t = \tilde{\Lambda}(\omega_\rho + \beta_\rho \rho_{t-1} + \alpha_\rho \times \frac{1}{q}\sum_{i=1}^{q}\Phi^{-1}(x_{t-i})\Phi^{-1}(y_{t-i})) \quad (7)$$

The connection function $\tilde{\Lambda}(\cdot)$ is defined as $\tilde{\Lambda}(\cdot) \equiv \frac{1-e^{-x}}{1+e^{-x}}$, which is set to ensure that $\rho_t$ is between (-1,1). Q is generally set to be less than or equal to 10.

Compared with Patton's time-varying model, the GAS model increases the flexibility of the parameter update mechanism and takes into account the inherent characteristics of the Copula function(Creal et al.(2013)), making it more robust(Yanlin,2022). The model assumes that $y_t \in R^N$ is an N-dimensional random vector at time t and $y_t = \mu + \varepsilon_t$, where $\mu$ represents the expected value of $y_t$ and $\varepsilon_t$ obeys a density function based on the observed value:

$$\varepsilon_t \sim p(\varepsilon_t|\theta_t, F_t\ ;\xi). \quad (8)$$

Among them, $F_t \equiv y_{1:t-1} \equiv (y_1', \ldots, y_{t-1}')'$ contains the past values of $y_t$ until time t-1, representing the information set at time t. $\theta_t \in \Theta \subseteq R^J$ is a time-varying parameter vector that controls p(·). $\theta_t$ depends only on $y_{1:t-1}$ and a set of static additional parameters $\xi$ that satisfy $\theta_t \equiv \theta(y_{1:t-1},\xi)$. The main feature of the GAS model is that the evolution of the time-varying parameter $\theta_t$ is driven by the conditional distribution score determined by (9), which has an autoregressive property:

$$\theta_{t+1} \equiv k + As_t + B\theta_t. \quad (9)$$

Where (k, A, B) = $\xi$ is a coefficient matrix with a specific dimension. In formula (9), $s_t$ is a vector proportional to the score of formula (10), $\nabla_t(y_t, \theta_t)$ is the score term, and the matrix $S_t$ is a J*J positive definite scaling matrix at time t. $I_t(\theta_t)$ is an information matrix related to the conditional distribution of $y_t$ given $y_{1:t-1}$, which is used to explain the change in $\nabla_t$.

$\gamma>0$, generally takes one of 0, 0.5, or 1.

$$s_t \equiv S_t(\theta_t)\nabla_t(y_t,\theta_t);$$

$$\nabla_t(y_t,\theta_t) \equiv \frac{\partial \log p(y_t;\theta_t)}{\partial \theta_t}; \quad (10)$$

$$S_t(\theta_t) \equiv I_t(\theta_t)^{-\gamma};$$

$$I_t(\theta_t) \equiv E_{t-1}\nabla_t(y_t,\theta_t)\,\nabla_t(y_t,\theta_t)';$$

In order to overcome the limitations of the original parameter range, parameter conversion is often necessary to meet the requirements of modeling. The standard processing method for the GAS model is to use a non-linear connecting equation $\Lambda(\cdot)$ as much as possible:

$$\theta_t = \Lambda(\tilde{\theta}_t) \quad (11)$$

$$\tilde{\theta}_t \equiv k + A\tilde{s}_t + B\tilde{\theta}_{t-1} \quad (12)$$

Among them, A and B are constant coefficients, $\tilde{s}_t \equiv \tilde{S}_t(\tilde{\theta}_t)\,\tilde{\nabla}_t(y_t,\tilde{\theta}_t)$, $\tilde{\nabla}_t(y_t,\tilde{\theta}_t)$ represents the score of formula (8) with respect to $\tilde{\theta}_t$. Referring to David et al. (2016), this article uses the mature GAS (1,1) model to dynamically modify the parameters of the binary Pair Copula function in the R-Vine tree structure.

2.4. Construction of Time-varying R-Vine Copula Model

The time-varying Vine Copula model is to make the constant parameters of the binary Copula and binary conditional Copula functions dynamic. The model based on the ARMA framework was the first to be applied to the dynamicization of Copula parameters. This paper uses the more excellent classification model GAS to dynamize Copula parameters instead. Different Copula function families have different distribution characteristics. This paper selects the Gussian Copula for fitting bivariate normal distribution, the Student-t Copula with double thick tail characteristics, the Gumbel Copula with upper tail characteristics, and the RotGumbel

Copula with lower tail characteristics to construct GAS-Copula. Therefore, the binary GAS-Copula constructed based on the above four types of Copula can meet the requirements of measuring different linkage characteristics between two variables.

The estimation of binary GAS-Copula parameters is currently mainly solved by maximizing the likelihood function (ML), and the solution formula:

$$\hat{\theta} \equiv \arg\max \log p(y_1; \theta_1) + \sum_{t=2}^{T} \log p(y_t; \theta_t). \tag{13}$$

Where $\theta_t$ is the parameter vector, $\theta_1 \equiv (I - B)^{-1}k$, B is the constant coefficient value in formula (9), and k is the constant intercept term in formula (9). $\theta_t \equiv \theta(y_{1:t-1}; \vartheta)$, $\vartheta \equiv (k, A, B)$, $p(y_t; \theta_t)$ is the likelihood probability function at time t. Based on the above discussion, this article chooses the R-Vine structure combined with the binary GAS-Copula to construct the time-varying Vine Copula model. During the modeling process, the R-Vine structure is more complex than the D-Vine and C-Vine structures in terms of the tree structure selection at each layer. For the same six variables, the C-Vine first layer tree has six possible structures, the D-Vine first layer tree has 6! structures, and the R-Vine structure has 32 * 6! structures. In the construction of the second and lower tree structures, especially D-Vine, the first tree directly determines the unique tree structure of other layers, and C-Vine is also relatively simple. However, the second and lower trees of R-Vine still need to be selected from multiple possible tree structures of the layer, which are often more complicated than the construction of the first tree structure. Therefore, the R-Vine structure inherently possesses greater detail than the previous two Vine structures, reflecting more realistic inter-variable dependencies. Applying the dynamic results of the GAS model parameters to solve for variables in the lower tree based on the upper tree is also a key modeling step. The detailed modeling process is described in the following steps and algorithm.

After the data obtained by filtering the raw data through the ARFIMA-GARCH model is extracted and converted into a uniformly distributed sequence on [0,1], the construction algorithm of the time-varying R-Vine Copula model can be summarized as follows:

(1) Following the principle that the edge of the next tree layer is determined by the two adjacent nodes of the previous tree layer, it is necessary to use uniformly distributed data values representing marginal distribution values or conditional marginal distribution values (for the second and lower trees) to calculate the absolute value of the Kendall's tau rank correlation coefficient between the two variables of the tree layer.

(2) The absolute values of Kendall's tau rank correlation coefficients of several possible non-closed-loop tree structures of this layer are summed up separately, and the tree structure with the smallest sum is the optimal structure of this layer;

(3) The GAS-Copula type between each pair of nodes in the tree structure of the determined layer is selected. The AIC criterion is used to select several GAS-Copulas after parameter estimation. The one with the smallest AIC value is the most suitable time-varying Pair-Copula. This article uses four time-varying Pair-Copulas, namely GAS-Gaussian-Copula, GAS-Student's t-Copula, GAS-Gumbel-Copula, and GAS-Rotgumbel-Copula, as the Pair Copulas between each pair of nodes in each layer of the tree.

(4) After the time-varying Pair-Copula type between two nodes in this layer of tree is determined, the conditional edge distribution value required by the next layer of tree can be calculated using dynamic parameters, and then the next layer of tree structure can be selected and the time-varying Pair-Copula type between two nodes can be determined;

(5) The calculation is repeated iteratively until the last layer of the tree. The AIC value of the entire model is obtained by summing up the AIC values obtained by modeling every two nodes in each layer of the tree.

The parameter estimation between the edges or nodes is mainly solved by the formula (13) in the above content, that is, the likelihood function maximization (ML) method.

2.5. Parameter Estimation of Time-varying R-Vine Copula Model

The dynamicization of copula models was primarily pioneered by researcher Patton(2001), who used the ARMA process to dynamize copula parameters. He also contributed the Dynamic Copula Toolbox, a Matlab package for dynamic copula models. This toolbox, now in version 3.0, is one of the few and relatively mature dynamic copula modeling toolkits available. However, this toolkit can only provide time-varying Copula models using ARMA-type models for parameter dynamics, and only supports time-varying Vine-Copula modeling of C-Vine and D-Vine structures. In addition, the toolkit requires pre-estimation of appropriate parameters to start modeling and solving. Therefore, this is one of the reasons why time-varying Vine Copula models are currently rarely used, and the existing applied structures are all C-Vine and D-Vine. In order to apply the R-Vine structure to the construction of the time-varying Vine Copula model, this paper constructs an algorithm program for the time-varying R-Vine Copula model. The parameters of the time-varying R-Vine Copula model are solved by maximizing the following likelihood function:

$$L(\theta_1^m, \ldots, \theta_n^m; \theta_{1,t}^c, \ldots, \theta_{n(n-1)/2,t}^c) =$$

$$\sum_{t=1}^{z} \sum_{d=1}^{n} \log f_d(y_{d,t}; \theta_d) + \sum_{t=1}^{z} \sum_{e=1}^{n(n-1)/2} \log c_e(y_{1,t}, y_{2,t}; \theta_{e,t}); \quad (14)$$

$$\theta_{e,t} \equiv \theta_e(y_{1:t}; k, A, B).$$

Among them, $\theta_{e,t}$ is the parameter vector, B is the constant coefficient value in formula (9), and k is the constant intercept term in formula (9). Other references can be made to formulas (8) to (13).

This paper uses the GAS-Copula, a superior approach to Patton's time-varying Copula, to construct a time-varying R-Vine Copula model. As even more advanced time series models are bound to emerge in the future, this article considers how to apply these new time series models with dynamic parameters to the dynamic construction of R-Vine Copula models. Therefore, this article analyzes the key procedures for applying these models to the construction of time-varying R-Vine Copula models, providing insights for the continued improvement of time-varying R-Vine Copula models.

| (a) R-Vine spanning tree algorithm. |
|---|
| Input: n variables $u_1, u_2, \ldots, u_n$ uniformly distributed on [0,1], $u_n$ is used as $N_i$(point) of the first layer tree |
| 1: Calculate E (edges) and w (edge weights) between all $N_i$ (points) to form a tree structure set G=(N,E,w). <br> 2: Search Initial point $N^*$: calculate $\widetilde{w_i} = \sum_{j=1}^{n} w_{ij}$ & $N^*$=argmax$\{\widetilde{w_i}, i = 2, \ldots, n\}$. <br> 3: Take $N^*$ as the initial point, $\widetilde{N} := \{N^*\}$ and $\widetilde{E}=\emptyset$. <br> 4: while $\widetilde{N} \neq N$ do <br> 5: Choose an edge with minimum w: e = $\{N_i, N_j\}$ & $N_i \in \widetilde{N}$ & $N_j$ not $\in \widetilde{N}$ (i≠ j) <br> 6: Let $\widetilde{N} := \widetilde{N} \cup \{N_j\}$ and $\widetilde{E} := \widetilde{E} \cup \{e\}$. <br> 7: end while |
| Output: Optimal tree structure $\widetilde{G}=(\widetilde{N}, \widetilde{E})$. |

The R-Vine spanning tree algorithm is used to select the tree structure for each layer in part (b). The structure selection range includes C-Vine and D-Vine. The algorithm in part (a) mainly refers to the algorithm program of Brechmann (2010).

(b) Algorithm for solving parameters of the time-varying R-Vine Copula model.

Input: n raw variables $u_1, u_2, \ldots, u_n$

1: Use (a) algorithm to calculate the optimal tree structure between $u_1, u_2, \ldots, u_n$, that is, the first-level tree structure $G_1=(N, E^1, w)$.

2: if ($\{u_i, u_j\} \in E^1) \cap (i \neq j)$

    For s in (1,...,k)

        MLE$\{ Gas-Copula^s(u_i, u_j) \}$      # s is the sth Gas-Copula

        Return $AIC^s$ and $\xi^s=(k^s, a^s, b^s)$     # MLE is the maximum likelihood algorithm

        forwhich AIC=min$\{AIC^s\}$

        $Gas-Copula_{i,j} = Gas-Copula^s(u_i, u_j)$

    End if

3: for d>1                                                              #$G_d=(N, E^d, w)$

    while ($E_i^d \cap E_j^d =1) \cap (i \neq j)$ do    #"1" means that $E_i^d$ and $E_j^d$ have the same conditional variable

    $F_m^d = h(E^d; \xi_m^{d-1})$              #"m" represents the number of conditional edge variables involved in the tree at this layer

    Return $F_m^d$                        # h(·) is the formula (15) in the paper

    Use algorithm (a) to calculate the optimal tree structure between $F_1^d, \ldots, F_m^d$

    Compute the second 2nd process of d-layer tree

    End for

Output: Time-varying R-Vine Copula tree structure, parameters, AIC values, etc..

After the model structure and parameters are determined through processes (a) and (b), the simulation process in part (c) can be carried out.

(c) Model simulation algorithm.

First: Use the Monte Carlo simulation method to simulate n independent $w_i$(i=1,…,n) that are uniformly distributed on [0,1].

Then，set:

$$\begin{aligned} x_1 &= w_1 \\ x_2 &= F_{2|1}^{-1}(w_2| x_1) \\ x_3 &= F_{3|1,2}^{-1}(w_3| x_1, x_2) \\ .. &= \ldots \\ x_n &= F_{3|1,\ldots,n-1}^{-1}(w_n| x_1,\ldots, x_{n-1}) \end{aligned}$$

Where $F^{-1}(\cdot)$ is the inverse function of formula (15), and $x_1, x_2, \ldots, x_n$ are the simulated data results.

(a), (b), and (c) illustrate modeling and simulation algorithms based on maximum likelihood estimation. Since (c) differs little from existing time-varying D-Vine Copula and C-Vine Copula simulation and forecasting processes, only its key principles and steps are presented.

For future new and better time series models that can be used for parameter dynamic modeling, the key step in its application lies in the third row of part 2 of Table 2 in algorithm (b), namely the MLE { $Gas - Copula^s(u_i, u_j)$ }process.

3. Applications and Comparison with Time-Varying D-Vine and C-Vine Copula

To further validate the rationality of our new model and its superiority over existing models, we use the time-varying R-Vine Copula model to model liquidity data for China and major Southeast Asian countries from 2018 to 2023, analyzing the overall liquidity risk profile in the region. We also use the time-varying D-Vine and C-Vine copula models for comparative modeling.

3.1. Data Source and Description

Liquidity risk is generally categorized as macro-fundamental liquidity risk, money market liquidity risk, and capital market liquidity risk. Money market liquidity risk primarily plays an intermediary role, thus combining the characteristics of both macro and micro liquidity. Meanwhile, the overnight interbank lending rate (IBO) serves as a money market price. Therefore, this paper selects the IBO as the raw data, and draws on the indicators constructed by Kyle and Obizhaeva(2007) to measure liquidity risk. This liquidity risk indicator is $Lr_t$=ln($L_t$)-ln($L_{t-1}$), where $L_t$ is the IBO of a country on day t, which can represent the country's money market interest rate level. The paper selects relevant data from six countries: Singapore, Thailand, Malaysia, Indonesia, China, and Vietnam. The data period is from January 1, 2018 to May 31, 2023, excluding dates when the six countries did not announce interest rates on the same day. The paper data has a total of 1,093 periods of original data. Given that the

Basel Committee stipulates that the number of VaR backtesting periods should be no less than 250, this article selects a total of 400 periods of data from June 11, 2021 to May 31, 2023 for VaR backtesting. The data used in this paper were manually collected by the author from the websites of official institutions in various countries. The Indonesian data comes from JIBOR and INdONIA (Overnight) of BANK INDONESIA, the Vietnamese data comes from VNIBOR (Overnight) of THE STATE BANK OF VIETNAM, the Chinese data comes from SHIBOR (Overnight) of CFETS, the Thai data comes from the Weighted Average Interest of BANK OF THAILAND, the Malaysian data comes from the Overnight Weighted Average Interest (Not OPR) of BANK NEGARA MALAYSIA, and the Singapore data comes from SIBOR_3M of the Monetary Authority of Singapore.

In order to analyze the characteristics of the liquidity risk indicator data representing the six countries, and also to determine the distribution of moving average, autocorrelation and other orders and random disturbance terms in the ARFIMA-GARCH original data modeling process. This paper first uses descriptive statistics on the raw data to obtain data characteristics such as skewness, kurtosis, and stationarity. Furthermore, the descriptive statistics are used to explain the selection of different binary GAS-Copula types in the modeling of time-varying R-Vine Copulas and to analyze the modeling and simulation results.

**Table 1 Descriptive statistics of the data.**

| $L\,n(\Delta\,\text{IBO})$ | Mean | SD | Skewness | Kurtosis | Ljung-Box | ArchTest |
|---|---|---|---|---|---|---|
| CN | 0.00036756 | 0.1563183 | -1.360899 | 12.32076 | 6.015e-07 | 1.404e-05 |
| THA | -0.00028244 | 0.02953950 | -0.6370554 | 103.8317 | 0.9461 | 1 |
| SG | -0.00091276 | 0.02.144682 | -0.9290962 | 58.54883 | < 2.2e-16 | 0.09976 |
| MY | -0.00003105 | 0.01731650 | 1.620310 | 36.17450 | 4.49e-11 | 0.05148 |
| VN | -0.00091877 | 0.1450001 | -2.135624 | 26.31551 | < 2.2e-16 | 1.206e-15 |
| IND | -0.00033563 | 0.02171113 | 0.04589069 | 131.5684 | < 2.2e-16 | < 2.2e-16 |

From the descriptive statistics, it can be seen that the average Ln (Δ IBO) of the six countries is mainly at the level of a few ten-thousandths, among which Malaysia has the smallest value, indicating that Malaysia's IBO fluctuation range is the smallest. Except for the right-skewed data for Malaysia and Indonesia, the rates of change of the logarithmic liquidity levels of the other four countries are all left-skewed. This feature will be reflected in the choice of binary GAS-Copula in the time-varying R-Vine modeling below. From the perspective of kurtosis, the data of the six countries are far greater than 3, which shows that the data of the six countries have serious fat tails. Judging from the results of the Ljung-Box test and the ArchTest test, the data of Thailand have insufficient lag correlation and no ARCH effect (autocorrelation of conditionally heteroskedastic series). Data from other countries all have varying degrees of lagged correlation and ARCH effects. Therefore, it is necessary to use the ARFIMA-GARCH-sstd model to model these data.

This paper considers the Ljung-Box, ArchTest and AIC information criterion to determine p, d, q, a and b in the ARFIMA(p,d,q)-GARCH(a,b) model, and obtains the models corresponding to the time series of liquidity risk level values of China, Thailand, Singapore, Malaysia, Vietnam and Indonesia, which are ARFIMA(1,0,2)-GARCH(1,1); ARFIMA(2,d,1)-GARCH(0,1);ARFIMA(2,d,2)-GARCH(1,1);ARFIMA(2,d,2)-GARCH(1,1); ARFIMA(1,d,2)-GARCH(1,1); ARFIMA(2,d,0)-GARCH(1,1). The raw data of the six countries were modeled using the ARFIMA(p,d,q)-GARCH(a,b)-sstd model, and the residual data of the six countries were extracted and converted into standardized residuals. The standardized residual sequence was then converted into data with a uniform distribution of [0,1] through probability integration.

3.2. Modeling and Parameter Estimation Results

The ARMA-GARCH model was used to filter the raw data and perform time-varying R-

Vine Copula algorithm modeling. The results showed that the relationships between the data representing six countries were D-Vine parallel relationships (D-Vine is a specialized R-Vine), which is significantly different from the R-Vine structure obtained in the following modeling and also differs from the D-Vine in Figure 5-(2), as shown in Figure 4. In addition, Table 2 shows the Likelihood of the modeling results of two models on the raw data of six countries. Except for Singapore, the LLH values of the other five countries are relatively close. The Likelihood of the Singapore data, which is modeled using the fractional difference method, is about 614 larger than the original model. This reflects Singapore's important position in the international financial market, as financial market variable data often has non-stationarity and long-term memory. Therefore, the ARFIMA-GARCH model incorporating fractional differencing in the paper performs better than the original model.

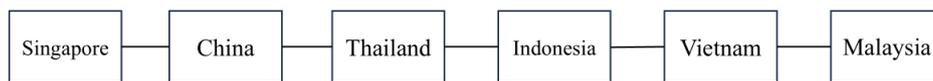

Figure 4. Modeling results of the R-Vine Copula program for ARMA-GARCH filtered data.

**Table 2 Log-likelihood results of raw data modeling for each country.**

|  | Model | CN | THA | IND | SG | MY | VN | Sum |
|---|---|---|---|---|---|---|---|---|
| Likelihood | ARMA-GARCH | -897.7 | -3335.2 | -4046.4 | -5644.7 | -3377.6 | -1180.0 | -18481.6 |
|  | ARFIMA-GARCH | -894.8 | -3379.1 | -4051.9 | -4980.4 | -3378.6 | -1182.4 | -17867.2 |

After using the ARFIMA-GARCH model to filter the raw data and extract the residual sequence, the residual sequence met the modeling requirements of the time-varying R-Vine Copula model. The modeling results in Table 3 show that the AIC values for the time-varying R-Vine Copula, time-varying D-Vine Copula, and time-varying C-Vine Copula modeling

**Table 3 Time-varying R-Vine/C-Vine/D-Vine Copula modeling and parameter estimation results.**

| Structure/Parameter | Edge/node1 | Edge/node2 | Edge/node3 | Edge/node4 | Edge/node5 |
|---|---|---|---|---|---|
| Tree1 | C1,4 | C4,3 | C3,2 | C3,6 | C6,5 |
|  | GAS-Rotgum | GAS-Rotgum | GAS-Gumbel | GAS-Rotgum | GAS-Rotgum |
|  | AIC=7.44 | AIC=8.60 | AIC=-0.94 | AIC=3.7017 | AIC=2.51 |
|  | $k=-0.1758$; | $k=-0.1936$; | $k=-0.2399$; | $k=-0.1642$; | $k=-0.1528$; |
|  | $a=0.0367$; | $a=-5.8603$; | $a=-0.173$; | $a=0.0777$; | $a=0.1353$; |
|  | $b=0.9589$. | $b=0.9579$. | $b=0.9248$. | $b=0.9582$. | $b=0.9566$. |
| Tree2 | C1,3\|4 | C2,4\|3 | C2,6\|3 | C3,5\|6 |  |
|  | GAS-Gumbel | GAS-Rotgum | GAS-Gumbel | GAS-Gumbe |  |
|  | AIC=4.48 | AIC=10.10 | AIC=1.0234 | AIC=6.9936 |  |
|  | $k=-0.181$; | $k=-0.2147$; $a=-0.0277$; | $k=-0.1474$; | $k=-0.1775$; |  |
|  | $a=0.0614$; | | $a=0.1244$; | $a=0.005$; |  |
|  | $b=0.9541$. | $b=0.9489$. | $b=0.9579$. | $b=0.9609$. |  |
| Tree3 | C1,2\|3,4 | C4,6\|2,3 | C2,5\|3,6 |  |  |
|  | GAS-Rotgum | GAS-Rotgum | GAS-Rotgum |  |  |
|  | AIC=4.15 | AIC=3.65 | AIC=7.2917 |  |  |
|  | $k=-0.1748$; | $k=-0.1393$; | $k=-0.1802$; |  |  |
|  | $a=0.0358$; | $a=-0.0803$; | $a=0.0053$; |  |  |
|  | $b=0.9568$. | $b=0.964$. | $b=0.9603$. |  |  |
| Tree4 | C1,6\|2,3,4 | C4,5\|2,3,6 |  |  |  |
|  | GAS-Rotgum | GAS-Rotgum |  |  |  |
|  | AIC=1.19 | AIC=7.10 |  |  |  |
|  | $k=-0.1679$; | $k=-0.1752$; |  |  |  |
|  | $a=0.2467$; | $a=0.0418$; |  |  |  |
|  | $b=0.9442$. | $b=0.9587$; |  |  |  |
| Tree5 | C1,5\|2,3,4,6 |  |  |  |  |
|  | GAS-Rotgum |  |  |  |  |
|  | AIC=7.17 |  |  |  |  |
|  | $k=-0.1752$; |  |  |  |  |
|  | $a=0.0179$; |  |  |  |  |
|  | $b=0.9604$. |  |  |  |  |
|  | time-varying R-Vine | | time-varying D-Vine | | time-varying C-Vine |
| **Sum(AIC)** | 72.4113 | | 72.4974 | | 80.6008 |

(**Note:** Cx,y|u,v represents an edge/node, x and y represent marginal distribution variables, and u and v are conditional variables. The binary GAS (1,1)-copula for each node was selected using the AIC criterion. k, a, and b are parameters of the GAS model. Due to space limitations, the complete results of the time-varying D-Vine/C-Vine copula are not presented. The number "1" represents the corresponding variable for China, the number "2" represents the corresponding variable for Thailand, the number "3" represents the corresponding variable for Indonesia, the number "4" represents the corresponding variable for Singapore, the number "5" represents the corresponding variable for Malaysia, and the number "6" represents the corresponding variable for Vietnam.)

results are 72.4113, 72.4974, and 80.6008, respectively. The AIC values indicate that the modeling results for time-varying R-Vine Copula are better than those for time-varying D-Vine Copula and C-Vine Copula, with C-Vine Copula having the worst modeling results. From Figures 5(3) and (2), we can see that the first-level tree structures of time-varying R-Vine Copula and time-varying D-Vine Copula are relatively similar, and their second-level trees and the tree structures below them are also very similar ( you can draw according to Table 3 and the tables in the appendix). This is the reason why their AIC values are very close. However, from Figure 5(1), we can see that the structures of time-varying C-Vine Copula and time-varying R-Vine Copula are very different, which is the reason for the large difference in their AIC values. This also shows that the time-varying C-Vine Copula structure fails to truly fit the relationship structure between the variables. If more variables were studied, the differences between the three would be more obvious.

Table 3 shows that the GAS-Copula for each edge/node in the five-layer tree is composed of a binary GAS-Gumbel-Copula and a binary GAS-Rotgumbel-Copula, with the binary GAS-Rotgumbel-Copula being the majority. This is consistent with the heavy-tailed and predominantly left-skewed characteristics of the data shown in the descriptive statistics in Table 1, indicating that the Gumbel-Copula focuses on measuring upper-tail correlations, while the Rotgumbel focuses on measuring lower-tail correlations.

From the first-level tree structure in Figure 5(3), we can see that the three ASEAN countries, Vietnam, Indonesia, and Singapore, occupy a central position in this regional liquidity risk spillover structure, with Indonesia in particular being at the core. This modeling result is firstly

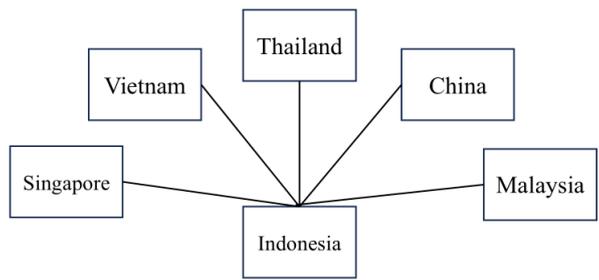

(1) First tree of Time-Varying C-Vine copula

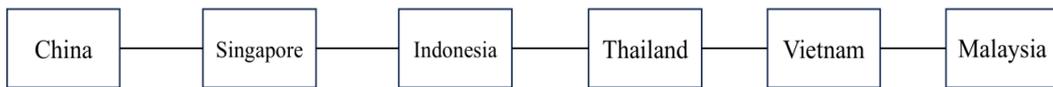

（2）First tree of Time-Varying D-Vine copula

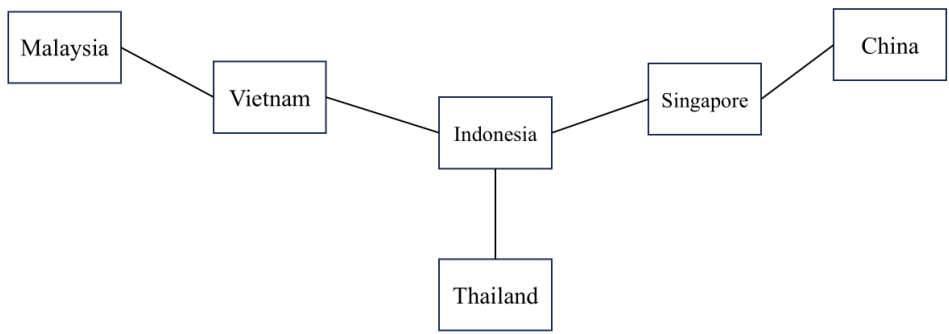

（3）First tree of Time-Varying R-Vine copula

Figure 5. The first-level tree of the time-varying C-Vine Copula, time-varying D-Vine Copula, and time-varying R-Vine Copula based on GAS dynamics.

related to Indonesia's geographical location. Indonesia borders Singapore, Malaysia, and Thailand, and its geographical proximity facilitates economic, trade, and financial exchanges. Indonesia's relatively mature financial system, high degree of marketization, and far-leading economic size among ASEAN countries give it a strong influence on other countries, which is reflected in the transmission of liquidity fluctuations. At the same time, the liquidity risk spillover structure centered on Southeast Asian countries also reflects the effectiveness of the ASEAN Community in building regional economic integration and deepens the linkage effect between the economic and financial systems of each other.

As can be seen from Figures 5(2) and (3), China and Singapore are closely connected in structure. The Singapore financial market is one of the most important international financial markets. A large number of Chinese companies are listed on the Singapore stock market, and Singapore is also one of the earliest countries to participate in Chinese investment activities. In addition, the Singapore money market is the center of the Asian dollar market, providing dollar financing and lending services to East Asian and Southeast Asian countries. The changes in the dollar in the Singapore money market and the changes in the dollar held by China, which has the largest dollar reserves, will inevitably affect each other, and thus affect the liquidity level of each country's domestic money market. These factors make the economic and financial ties between China and Singapore more frequent than those of other ASEAN countries, so the spillover relationship of liquidity risk between the two sides is also relatively close.

These economic, financial, and political realities not only reflect the objectivity of the R-Vine structure in measuring the region's liquidity risk status compared to the C/D-Vine structure, but also further demonstrate the success of the time-varying R-Vine Copula model constructed in this paper, enabling it to achieve better forecasting results than the latter two.

3.3. Simulation

After the time-varying R-Vine Copula model is used to model the data and determine the relevant parameters, the VaR value can be simulated and calculated. For the liquidity risk level value variables $L_1$、$L_2$、$L_3$、$L_4$、$L_5$, and $L_6$ of the six countries, the same weight (1/6) is assigned to each variable to construct the overall liquidity risk level value L=(1/6)*$L_1$ +(1/6)*$L_2$ +(1/6)*$L_3$ +(1/6)*$L_4$ +(1/6)*$L_5$ +(1/6)*$L_6$. Assuming the probability distribution function of L is F(L), then the VaR value of the indicator can be constructed: from Pr{L> δ}=$\int_{L>\delta}^{\cdot} d\,F(L)$=1-$\alpha$, we can get Pr{L≤ $VaR_\alpha$}=$\alpha$, where $\alpha$ is the confidence level value, and the confidence level $\alpha$ indicates that the probability that the overall liquidity risk indicator value is less than $VaR_\alpha$ is $\alpha$. The calculation of VaR value is divided into parametric method, semi-parametric method and non-parametric method. The main ones used are the historical simulation method of the semi-parametric method and the Monte Carlo simulation method of the non-parametric method. The article chooses Monte Carlo simulation method to obtain simulation data. Confidence levels of 0.9, 0.95, 0.99, and 0.995 are selected to simulate 1000 times at each time point t to obtain the VaR values of $VaR_t^{0.9}, VaR_t^{0.95}, VaR_t^{0.99}$, and $VaR_t^{0.99}$, and VaR ^ 0.99 at each time point t. These VaR values are located at the higher right end of the 1000 simulated values sorted by size. The key steps of simulating the six-country data using the time-varying R-Vine Copula model are obtained through multiple iterations of the inverse function of formula (15).

$$h(x,v)=F(x|v)= \frac{\partial C_{x,v_j|v_{-j}}(F(x|v_{-j}),F(v_j|v_{-j}))}{\partial F(v_j|v_{-j})} \ . \qquad (15)$$

The entire simulation process first undergoes Monte Carlo simulation to obtain six variable sequences that follow a uniform distribution on [0,1], and then undergoes inverse empirical probability integration to obtain a standardized residual sequence. After transforming the standardized residual sequence into a residual sequence, the simulated raw data, namely the logarithmic liquidity level change rate, is obtained through the inverse ARFIMA-GARCH process. Finally, the simulated raw data is used to calculate the VaR values for each confidence level. This simulation process is basically the inverse modeling process of the time-varying R-Vine Copula model. See the algorithm (c), and the details will not be repeated.

Figure 6 is a line chart showing the calculation results of the relevant indicator values. The horizontal axis is the time interval and the vertical axis is the logarithmic change value of the liquidity level. It can be seen from the figure that as the confidence level increases, the simulated VaR value continues to increase and the increase is expanding. The four broken lines from bottom to top represent the true liquidity risk level (black), the VaR value at a 90% confidence level (green), the VaR value at a 95% confidence level (red), and the VaR value at a 99.5% confidence level (purple). The true liquidity risk level reaches a maximum value exceeding 0.2, while the VaR value at a 99.5% confidence level exceeds 0.4 at multiple points in time. This indicates that the fluctuation trends of the VaR values at these three different confidence levels are generally consistent. From the time trend, from the beginning of 2020 to the beginning of 2021, the fluctuation range of the real liquidity risk level increased significantly and was at the highest level in the entire time period. The maximum value of the real value of the liquidity risk level also appeared during this period, and the VaR values of the three confidence levels also showed a concentrated surge. With the outbreak of the COVID-19 pandemic in early 2020, China and ASEAN countries adopted strict prevention and control measures. These prevention and control measures have put the social and economic operations in many regions and time periods into a "silent" state. The operating difficulties of various economic entities have led to the depletion of cash flow, which in turn has put the liquidity of the entire country into a tight state. Various liquidity risk events also broke out during this

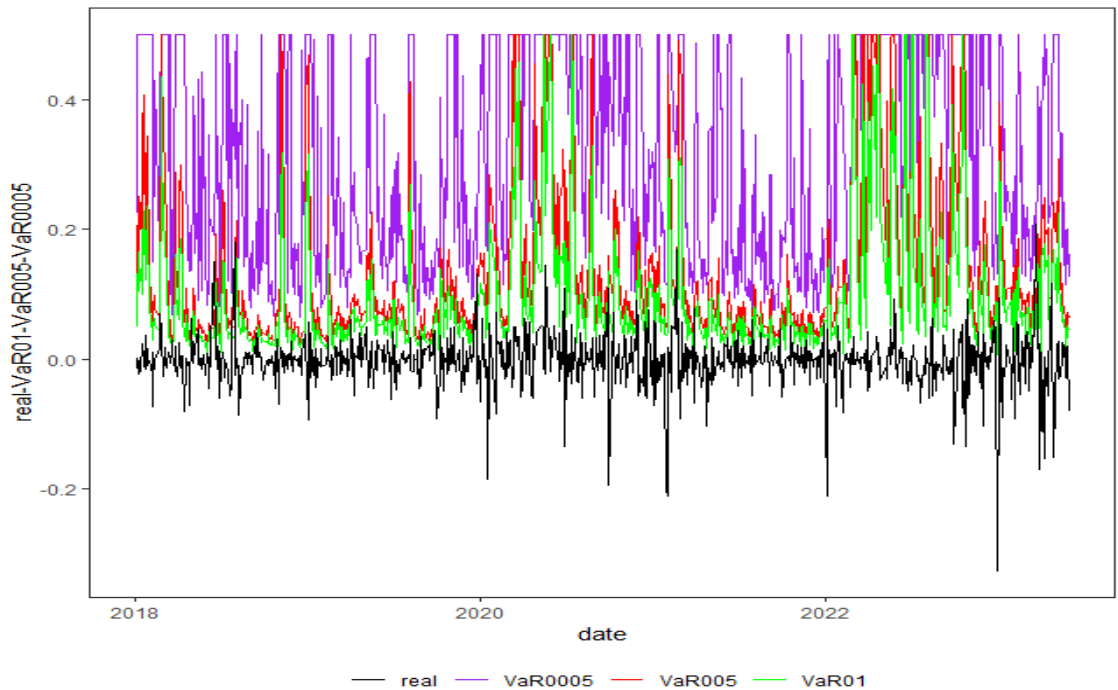

Figure 6 Prediction results of liquidity risk level in this region

(Note: For a clearer graphical display, the VaR line chart for α=0.99 is not shown, see Appendix)

period.Therefore, this modeling result reflects the actual liquidity risk situation during this period, validating the reliability of the model's predictions. Starting in early 2022, the actual fluctuations in liquidity levels shifted from flat to gradually increasing, reaching their peak level by the end of the year. These fluctuations can be explained by the following events. On January 1, 2022, the RCEP agreement entered into force between China and major Southeast Asian countries, further facilitating economic, trade, and financial exchanges between the two countries and alleviating liquidity risks. In the second and third quarters of 2022, China's economy experienced a severe downturn. Rising international food and energy prices posed a risk of stagflation to the global economy, further exacerbating liquidity risks in the region. Therefore, these economic and financial factors, both within and outside the region, jointly shaped the liquidity risk situation after early 2022.

**Table 4 Kupiec test results of VaR values of time-varying R-Vine Copula**

| $\alpha$ | Fail times | Fail rate | P-value | LR-value | Loss | MAD |
|---|---|---|---|---|---|---|
| 90% | 14 | 3.5% | 7.1e-07 | 24.6 | 0.0137 | 1.0998 |
| 95% | 8 | 2.0% | 0.0018 | 9.8 | 0.0103 | 1.6493 |
| 99% | 2 | 0.5% | 0.2640 | 1.2 | 0.0088 | 14.0323 |
| 99.5% | 2 | 0.5% | 0.9972 | 1.25e-05 | 0.0108 | 48.2385 |
| Real v | Min=-0.3268; | | Max=0.2074; | | Mean=-0.00035; | |

Table 4 shows the Kupiec test of VaR values at four different confidence levels. This test is based on the binomial distribution theory and uses the likelihood ratio to evaluate the accuracy of the model by comparing the differences between the simulated expected events and the actual events. Among them, the likelihood ratio LR value is $LR=-2*\log(((1-\alpha)^{\wedge}(T-N))*(\alpha^{\wedge}N))+2*\log(((1-N/T)^{\wedge}(T-N))*((N/T)^{\wedge}N))$, N is the number of failures and T is the total number of failures. The last row shows the minimum, maximum, and mean of the true value of regional liquidity risk. From the results of the Kupiec test, the prediction effects of α of 99% and 99.5% are better, among which the failure rate of 99.5% is closest to the value of 1-α. The VaR values with α of 90% and 95% failed the Kupiec test, indicating that the VaR values at these two confidence levels overestimate the risk to a certain extent. This may be due to Vietnam's larger |Ln (Δ IBO)| among the six countries, which affects the overall forecast accuracy. This reflects the prudence of the risk measurement model.

Tables 5 and 6 are the Kupiec test results of VaR values simulated by the time-varying D-Vine Copula model and the time-varying C-Vine Copula model respectively. Combined with the test results in Table 4, it can be seen that when α is 90% and 95%, both time-varying D-

**Table 5 Kupiec test results of VaR values of time-varying D-Vine Copula**

| α | Fail times | Fail rate | P-value | LR-value | Loss | MAD |
|---|---|---|---|---|---|---|
| 90% | 15 | 3.75% | 2.366e-06 | 22.27 | 0.0136 | 0.9974 |
| 95% | 5 | 1.25% | 4.329e-05 | 16.72 | 0.0103 | 1.5775 |
| 99% | 2 | 0.50% | 0.2660 | 1.24 | 0.0097 | 16.2023 |
| 99.5% | 1 | 0.25% | 0.4325 | 0.62 | 0.0109 | 48.5048 |
| Real v | Min=-0.3268; | | Max=0.2074; | | Mean=-0.00035; | |

**Table 6 Kupiec test results of VaR values of time-varying C-Vine Copula**

| $\alpha$ | Fail times | Fail rate | P-value | LR-value | Loss | MAD |
|---|---|---|---|---|---|---|
| 90% | 15 | 3.75% | 2.358e-06 | 22.27 | 0.0135 | 1.0404 |
| 95% | 8 | 2.00% | 0.0018 | 9.71 | 0.0102 | 2.0222 |
| 99% | 2 | 0.50% | 0.2660 | 1.24 | 0.0098 | 19.2852 |
| 99.5% | 1 | 0.25% | 0.4325 | 0.62 | 0.0113 | 55.9211 |
| Real v | Min=-0.3268; | | Max=0.2074; | | Mean=-0.00035; | |

Vine and time-varying C-vine fail the Kupiec test, and the difference between time-varying R-Vine and the two is small. The Kupiec test results at the 99% and 99.5% levels show that both passed the test, but at these two confidence levels, especially when α is 99.5%, the time-varying R-Vine's prediction effect on extreme liquidity risk events is significantly better than that of the time-varying D-Vine Copula and time-varying C-Vine Copula. Therefore, the Kupiec test results in Tables 4, 5, and 6 further verify that the modeling effect of the time-varying R-Vine Copula model constructed in this article is better than that of the time-varying D-Vine Copula and time-varying C-Vine Copula models.

3.4. Robustness Test of Simulation

In systemic financial risk events, different countries experience varying impacts in the risk spillover chain. Liquidity risk primarily spreads regionally through trade, investment, and external financial channels. A country's overall economic strength fundamentally determines its import and export volume, outbound investment and foreign investment attraction, and international financial standing. Therefore, this paper reassigned weights to the six countries based on their GDP. The GDP of the six countries was aggregated over the five years from 2018 to 2022, and the weight of each country was calculated as the proportion of its combined GDP over the five-year period to the total GDP of the six countries over the same period. The data comes from the World Bank database, and the results of the Kupiec test on the VaR values calculated using these weights are shown in Table 7.

**Table 7 Robustness test - Kupiec test results**

| $\alpha$ | Fail times | Fail rate | P-value | LR-value | Loss | MAD |
|---|---|---|---|---|---|---|
| 90% | 37 | 9.25% | 0.6016 | 0.273 | 0.0322 | 1.6827 |
| 95% | 21 | 5.25% | 0.8289 | 0.047 | 0.0210 | 1.8155 |
| 99% | 5 | 1.25% | 0.6323 | 0.356 | 0.0071 | 2.6361 |
| 99.5% | 3 | 0.75% | 0.7883 | 0.476 | 0.0044 | 6.2420 |
| Real v | Min=-1.2760; | | Max=0.5621; | | Mean=-0.00025; | |

The Kupiec test results in Table 7, including the P-value and the proximity of the failure rate to 1-$\alpha$, show that the VaR values at all $\alpha$ levels pass the test. This demonstrates the robustness of the time-varying R-Vine Copula model and its ability to accurately predict extreme liquidity risk events between China and Southeast Asian countries, further demonstrating the success of the model constructed in this paper.

## 4. Conclusion

This paper constructs a time-varying R-Vine Copula model based on the R-Vine structure algorithm and uses the GAS model for dynamic model parameter optimization. This model not only takes the C-Vine/D-Vine structure into consideration when modeling, but also incorporates the R-Vine structure that is not a C-Vine or D-Vine structure into the range of structural selection for the construction of the time-varying Vine Copula model. In empirical applications, the modeling results of this model are not only better than those of the time-varying C-Vine/D-Vine Copula models in a series of statistical tests, but in the study of liquidity risk spillover issues between Southeast Asian countries and China, the time-varying R-Vine Copula model also reflects the economic, financial and even political relations in the region more realistically than the latter two models. Therefore, the paper concludes that the R-Vine structure is the optimal choice for time-varying Vine Copula models in economics or business research. The paper also finds that the effectiveness of modeling the residuals of the raw marginal distribution data influences the choice of Vine structure. Finally, the paper proposes a time-varying R-Vine Copula algorithm, providing insights for the application of new models with dynamic parameters in the construction of time-varying R-Vine Copula models.

Appendix:

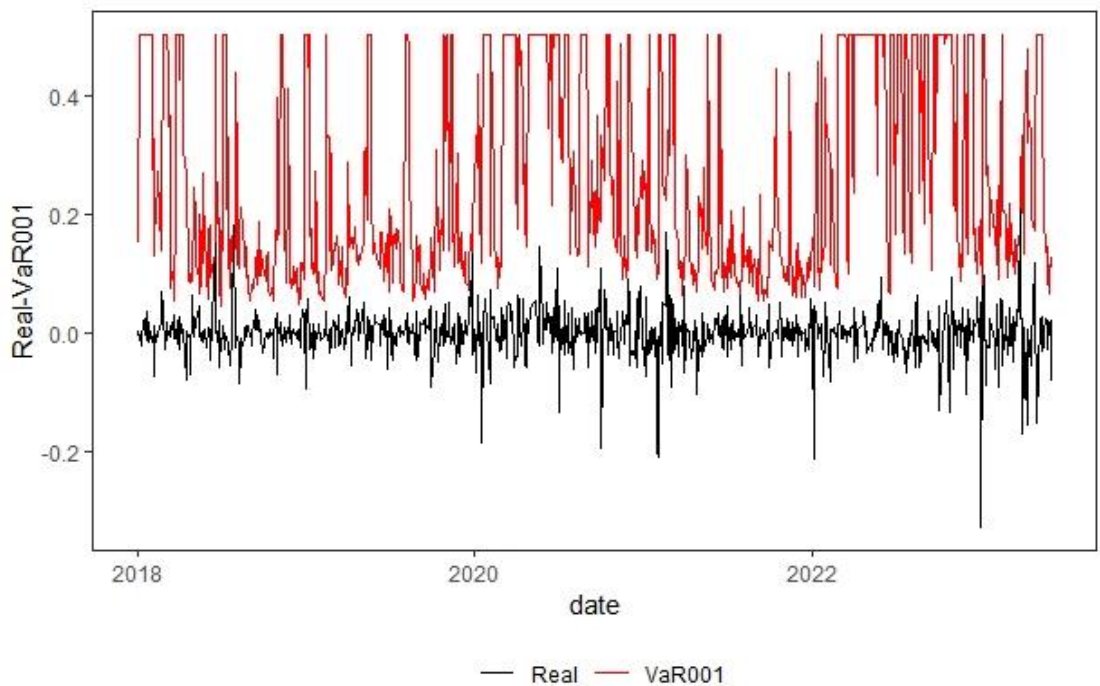

Figure　Line graph of true value of liquidity risk and VaR with α=0.99.

**TableA Time-varying C-Vine Copula modeling and parameter estimation results**

| Structure/ Parameter | Edge/node1 | Edge/node2 | Edge/node3 | Edge/node4 | Edge/node5 |
|---|---|---|---|---|---|
| Tree1 | C1,3<br>GAS-Rotgum<br>AIC=5.3006<br>k=-0.1840;<br>a=0.0474;<br>b=0.9549. | C2,3<br>GAS-Gumbe<br>AIC=-0.9408<br>k=-0.2399;<br>a=-0.173;<br>b=0.9248. | C4,3<br>GAS- Rotgum<br>AIC=8.5912<br>k=-0.1936 ;<br>a=-5.8603;<br>b=0.9579. | C5,3<br>GAS-Rotgum<br>AIC=6.3207<br>k=-0.2029;<br>a=-0.0281;<br>b=0.9515. | C6,3<br>GAS-Rotgum<br>AIC=3.7017<br>k=--0.1642 ;<br>a=0.0777;<br>b=0.9582. |
| Tree2 | C2,4\|3<br>GAS- Rotgum<br>AIC=10.1042<br>k=-0.2147;<br>a=-0.0277;<br>b=0.9489. | C6,4\|3<br>GAS-Rotgum<br>AIC=3.8431<br>k=-0.2074;<br>a=-0.0703;<br>b=0.9468. | C1,4\|3<br>GAS- Rotgum<br>AIC=7.3589<br>k=-0.1752;<br>a=0.0390;<br>b=0.9589. | C5,4\|3<br>GAS- Rotgum<br>AIC=7.4944<br>k=-0.1811;<br>a=0.0403;<br>b=0.9568. | |
| Tree3 | C1,2\|3,4<br>GAS-Rotgum<br>AIC=4.1541<br>k=-0.1748;<br>a=0.0358;<br>b=0.9568 . | C6,2\|3,4<br>GAS-Rotgum<br>AIC=3.4023<br>k=-0.1696;<br>a=0.0895;<br>b=0.9543. | C5,2\|3,4<br>GAS-Rotgum<br>AIC=7.3113<br>k=-0.1807 ;<br>a=0.0048;<br>b=0.9603. | | |
| Tree4 | C1,6\|2,3,4<br>GAS-Rotgum<br>AIC=1.1895<br>k=-0.1679;<br>a=0.2467;<br>b= 0.9442. | C5,6\|2,3,4<br>GAS-Rotgum<br>AIC=5.6017<br>k=-0.1622 ;<br>a=0.0339;<br>b=0.9612; | | | |
| Tree5 | C1,5\|2,3,4,6<br>GAS-Rotgum<br>AIC=7.1679<br>k=-0.1752;<br>a=0.0179;<br>b =0.9604. | | | | |
| **Sum(AIC)** | | 80.6008 | | | |

The number "1" represents the corresponding variable for China, the number "2" represents the corresponding variable for Thailand, the number "3" represents the corresponding variable for Indonesia, the number "4" represents the corresponding variable for Singapore, the number "5" represents the corresponding variable for Malaysia, and the number "6" represents the corresponding variable for Vietnam.

**TableB Time-varying D-Vine Copula modeling and parameter estimation results**

| Structure/Parameter | Edge/node1 | Edge/node2 | Edge/node3 | Edge/node4 | Edge/node5 |
|---|---|---|---|---|---|
| Tree1 | C1,4<br>GAS-Rotgum<br>AIC=7.44<br>k=-0.1758;<br>a=0.0367;<br>b=0.9589. | C4,3<br>GAS-Rotgum<br>AIC=8.60<br>k=-0.1936;<br>a=-5.8603;<br>b=0.9579. | C3,2<br>GAS-Gumbel<br>AIC=-0.94<br>k=-0.2399 ;<br>a=-0.173;<br>b=0.9248. | C2,6<br>GAS- Gumbel<br>AIC=-0.0767<br>k=-0.14531;<br>a=0.122608;<br>b=0.9586. | C6,5<br>GAS-Rotgum<br>AIC=2.51<br>k=-0.1528 ;<br>a=0.1353;<br>b=0.9566. |
| Tree2 | C1,3\|4<br>GAS-Gumbel<br>AIC=4.48<br>k=-0.181;<br>a=0.0614;<br>b=0.9541. | C2,4\|3<br>GAS-Rotgum<br>AIC=10.10<br>k=-0.2147; a=-0.0277;<br>b=0.9489. | C3,6\|2<br>GAS- Rotgum<br>AIC=3.7115<br>k=-0.1683;<br>a=0.0881;<br>b=0.9561. | C2,5\|6<br>GAS- Rotgum<br>AIC=7.2610<br>k=-0.1801;<br>a=0.0057;<br>b=0.9603. | |
| Tree3 | C1,2\|3,4<br>GAS-Rotgum<br>AIC=4.15<br>k=-0.1748;<br>a=0.0358;<br>b=0.9568 . | C4,6\|2,3<br>GAS-Rotgum<br>AIC=3.65<br>k=-0.1393;<br>a=-0.0803;<br>b= 0.964. | C3,5\|2,6<br>GAS-Rotgum<br>AIC=6.1539<br>k=-0.1954 ;<br>a=-0.0154;<br>b=0.9550. | | |
| Tree4 | C1,6\|2,3,4<br>GAS-Rotgum<br>AIC=1.19<br>k=-0.1679;<br>a=0.2467;<br>b= 0.9442. | C4,5\|2,3,6<br>GAS-Rotgum<br>AIC=7.10<br>k=-0.1752 ;<br>a=0.0418;<br>b=0.9587; | | | |
| Tree5 | C1,5\|2,3,4,6<br>GAS-Rotgum<br>AIC=7.17<br>k=-0.1752;<br>a=0.0179;<br>b =0.9604. | | | | |
| **Sum(AIC)** | | **72.4974** | | | |

## Disclosure Statement

The authors report there are no competing interests to declare.